**Spin Relaxation in Weak Localization Regime in Multilayer Graphene Spin Valves**

Takehiro Yamaguchi[1,*], Rai Moriya[1,*], Satoru Masubuchi[1,2], Kazuyuki Iguchi[1], and Tomoki Machida[1,2,*]

[1] *Institute of Industrial Science, University of Tokyo, Meguro, Tokyo 153-8505, Japan*

[2] *Institute for Nano Quantum Information Electronics, University of Tokyo, Meguro, Tokyo 153-8505, Japan*

[3] *PRESTO, Japan Science and Technology Agency, Kawaguchi, Saitama 332-0012, Japan*

The temperature dependence of the spin relaxation time $\tau_s$ in multilayer graphene (MLG) spin valve devices was measured using a non-local magnetoresistance (NLMR) measurement. A weak localization (WL) was observed from magnetoresistance (MR) measurements below ~70 K, suggesting coherent transport of the charge carriers. Within the same temperature range, we observed a large increase in the spin relaxation time $\tau_s$ and spin diffusion length $\lambda_s$ even though the diffusion constant $D_s$ was suppressed by the WL. This demonstrated that the spin relaxation time in MLG could be significantly extended when the charge experiences quantum interference effect in the coherent charge transport regime.

E-mail: yamatake@iis.u-tokyo.ac.jp; moriyar@iis.u-tokyo.ac.jp; tmachida@iis.u-tokyo.ac.jp

Due to its weak spin-orbit and hyperfine interactions, graphene has recently received much attention for its potential spintronics applications[1]. Both electrical spin injection from a ferromagnet to graphene and electrical detection of spin transport have already been successfully demonstrated by several groups[2-6]. At the same time, charge transport in graphene is known to have a long phase coherence length. Quantum interference phenomena such as weak localization (WL) and Fabry-Perot interference have been observed in single-, bi-, and multi-layer graphene devices[7-10]. Therefore, graphene is an ideal material for studying spin transport in the coherent transport regime. Such transport properties can be thoroughly investigated by using the electrical spin injection technique, which is valid for few material systems. In this letter, we report the measurement of the temperature dependence of magnetoresistance (MR) and non-local magnetoresistance (NLMR) in multilayer graphene (MLG) spin valve devices. Comparing the two measurements, we find that the spin relaxation time is increased dramatically when the charge experiences the quantum interference effect in the coherent transport regime.

In these experiments, we characterized multilayer graphene spin valve devices. The typical data obtained from a single device are presented in this paper. MLG has various advantages that make it suitable for this experiment: 1) It exhibits a longer spin diffusion length compared to single and bi-layer graphene[11], and 2) at the same time, MLG displays a much more pronounced WL than single-layer graphene [12].

A flake of MLG was transferred onto a $SiO_2$/Si substrate with an oxide thickness of 300 nm by the micromechanical cleavage of Kish graphite. The device structure of the fabricated device is schematically shown in Fig. 1(a). Ferromagnetic permalloy (Py, $Ni_{81}Fe_{19}$) electrodes and nonmagnetic Au/Ti electrodes were fabricated on a MLG flake using conventional EB-lithography (Elionix ELS-7500) and EB-evaporation techniques. Two adjacent Py electrodes are designed to have different widths of 100 and 300 nm, respectively, in order to introduce different coercive fields. The separation, $L$, and channel width, $W$, between the Py electrodes are $L = 5.5$ μm and $W = 1.5$ μm, respectively. The layer number of the MLG is $7 \pm 1$, as determined from atomic force microscope measurements. To alleviate spin absorption at the ferromagnetic electrode[13], we introduced an $Al_2O_3$ tunnel barrier grown by atomic layer deposition (ALD) between the ferromagnetic electrodes and graphene[14]. The number of ALD cycles is six, and the obtained thickness of the $Al_2O_3$ is 0.6−0.8 nm. The contact resistance $R_c$ between the Py and graphene is on the order of 1−10 kΩ. The transport measurements were performed in an atmosphere of helium gas at 2−300 K using the standard lock-in technique.

First, we characterized the resistance of MLG as a function of the back-gate voltage $V_{BG}$. The contact geometry used for the measurement is shown in Fig. 1(a). We used an ac current amplitude of 1 μA for this measurement. The charge neutrality point (CNP) was placed outside our maximum voltage range of $V_{BG}$ = -60 to +60 V. The shift of the CNP was due to the hole doping from charged impurities existing on both the top and bottom surfaces of the MLG[15-17]. The hole density was estimated to be $1.1 \times 10^{13}$ cm$^{-2}$ at 300 K and for $V_{BG} = 0$ V. The carrier mobility $\mu = \left(R_{sq.} en\right)^{-1}$ was 1200 cm$^2$V$^{-1}$s$^{-1}$ at $V_{BG} =$ 0 V[18], comparable to the value observed in the MLG spin valve[11]. Next, we characterized the dependence of the MR on the out-of-plane magnetic field for different values of temperature. At low temperature, a negative MR was observed in the low-magnetic-field region [Fig. 1(b)]. We found that this occurred due to the WL of the MLG[7-9]. The MLG has small spin-orbit coupling, and therefore, WL is naturally observable in the MLG samples when the phase coherence time $\tau_\varphi$ exceeds the elastic scattering time. The height of the WL peak $\Delta\rho_{WL}$ is dominated by the $\tau_\varphi$ according to the theory of quantum interference in graphene[19]. The variation in the resistance of the MLG $R_{xx}$ at a zero magnetic field and the change in the height of the WL peak $\Delta\rho_{WL}$ for different temperatures are shown in Figs. 1(c) and 1(d), respectively. Both plots show a rapid increase below ~70 K. In the case of graphene, the dominant dephasing mechanism at a high temperature is electron-electron (*e-e*) interaction[7]. Thus, the appearance of WL indicates suppression of the *e-e* interaction at low temperature. We also observed that an oscillating variation in the resistance is superimposed on the MR curve in Fig. 1(b). This is attributed to the universal conductance fluctuation that arises from the quantum interference [7-9]; this provides another piece of evidence for the phase coherent transport of MLG at low temperature.

We characterized the spin transport properties of the MLG device using an NLMR measurement. The contact geometry for the measurement is schematically shown in Fig. 2(a). The NLMR ($R_{NL} = V_{NL}/I$) as a function of the in-plane magnetic field at 300 K and for $V_{BG} = 0$ V is shown in Fig. 2(b). When the magnetizations of the two Py electrodes are anti-parallel, an abrupt resistance change of $\Delta R_{NL} = 15$ m$\Omega$ due to the spin injection and spin transport in graphene is observed. We think that the contact resistance between the Py and graphene is in the so-called intermediate regime[13,14]. The calculated ratio between the contact resistance $R_c$ and the spin resistance of graphene $R_G = \lambda_s R_{sq}/W$ is $R_c/R_G \sim 2.1$ at 300 K, where $\lambda_s$ is the spin diffusion length, $R_{sq}$ is the sheet resistance of the MLG, and *W* is the width of the MLG. Therefore, the spin absorption of the ferromagnetic electrode is reasonably well suppressed.

The Hanle effect is measured under the out-of-plane magnetic field. We analyzed the obtained Hanle effect using the following equation[20]:

$$R_{NL} \propto \int_0^\infty \frac{1}{\sqrt{4\pi D_s t}} exp\left(-\frac{L^2}{4D_s t}\right) \cos\left(\frac{g\mu_B B t}{\hbar}\right) exp\left(-\frac{t}{\tau_s}\right) dt, \quad (1)$$

where $D_s$ is the spin diffusion constant, $\tau_s$ is the spin relaxation time, $L$ is the distance between the Py electrodes, $g$ is the electron $g$-factor, and $\mu_B$ is the Bohr magneton. The measured Hanle effect and the best-obtained fit to Eq. (1) at various temperatures are plotted in Fig. 2(c). The Hanle effect measured in both the parallel (*P*) and anti-parallel (*AP*) magnetic configurations are plotted together. The experimental data and the fitted curve strongly coincide. The extracted temperature dependences of $D_s$, $\tau_s$, $\lambda_s$, and $\Delta R_{NL}$ are shown in Fig. 2(d). Note that the extracted $D_s$, $\tau_s$, and $\lambda_s$ are slightly different in the *P* and *AP* configurations of magnetization. Therefore, we present in Fig. 2(d) the values obtained by averaging the results from the *P* and *AP* configurations.

Significant changes in the magnitudes of various parameters can be seen in Fig. 2(d) at temperatures below ~70 K. Notably, below this temperature, WL is observed in the MR measurement. With decreasing temperature, the spin diffusion constant $D_s$ tends to saturate and then decreases. We think that the suppression of $D_s$ can be attributed to the WL, since the conductance of charge is suppressed in the WL regime. However, the spin relaxation time $\tau_s$ significantly increases below this temperature, and as a result, even though $D_s$ decreases, the spin diffusion length $\lambda_s$ increases. A spin diffusion length $\lambda_s$ of 8 μm is determined. Remarkably, the increase in $\tau_s$ and $\lambda_s$ coincides with the emergence of WL [Fig. 1(d)]. This suggests that there is a correlation between these two phenomena. Below ~70 K, $R_{NL}$ saturates and then decreases with decreasing temperature. The reduction in $\Delta R_{NL}$ could be attributed to the increase in the spin resistance of the graphene. The spin resistance of the graphene $R_G$ ($\propto \lambda_s$) increases by a factor of almost 2 below ~70 K according to the change in the spin diffusion length. At the same time, $R_c$ increases by only ~3% from room temperature to 2 K. This decreases the ratio $R_c/R_G$ significantly, and as a result, $\Delta R_{NL}$ decreases because of the larger spin absorption effect at the Py/MLG interface.

Presently, the origin of the large increase in $\tau_s$ and $\lambda_s$ at low temperature remains unclear. Therefore, we offer some possible explanations for our observations. First, in graphene, there are two dominant spin relaxation mechanisms: Elliott-Yafet (EY) and Dyakonov-Perel (DP). In these two spin relaxation mechanisms, the inverse of the spin relaxation time $1/\tau_s$ is either directly proportional or inversely proportional to the charge diffusion constant $D_c$. If we assume $D_c = D_s$, then EY appears to be the dominant mechanism in our MLG sample at temperatures above ~70 K, where our measured data follow the relation $1/\tau_s \propto 1/D_c$. However, the assumption of neither of the spin relaxation mechanisms can clarify our experimental results for temperatures below ~70 K, since the change in $\tau_s$ is considerably rapid compared to that of $D_c$. The contact resistance of the Py is in the intermediate regime in our device. Recently, the effect of the intermediate contact on $\tau_s$ has been discussed by another group[21]. However, this contribution gives rise to a decrease in the spin relaxation time rather than an increase. Thus, the effect of the contact is not sufficient to explain the results. From our results, we think that there is a correlation between the spin relaxation time and the emergence of WL. Since the emergence of WL is related to the suppression of the *e-e* interaction, our results might suggest the importance of the *e-e* interaction in the spin transport of MLG. However, WL itself could influence the spin relaxation time. The correlation between WL and spin transport has previously been discussed for two-dimensional semiconductor systems [22,23]. These studies revealed that the constructive interference of the electron wavefunction under the WL could give rise to an enhancement in the spin relaxation. The effect of quantum interference on the coherent spin transport in graphene-based devices might need to be considered.

In summary, we performed a detailed comparison between the temperature dependences of the non-local spin signal $\Delta R_{NL}$, the Hanle effect, and the magnetoresistance in a multilayer graphene spin valve device. We demonstrated that the spin relaxation time and spin diffusion length are dramatically increased by the quantum interference effect in MLG.

Figure captions

**Fig. 1.** (Color online) (a) Schematic illustration of the device and MR measurement configuration. An ALD-Al$_2$O$_3$/PTCA tunnel barrier was introduced between the MLG and the electrodes. (b) MR measured under out-of-plane magnetic field $B_\perp$ at different temperatures, showing WL features. The traces are offset for clarity. Also shown are the temperature dependence of (c) the resistance of graphene at zero magnetic field and (d) the height of the WL peak $\Delta\rho_{WL}$.

**Fig. 2.** (Color online) (a) Schematic illustration of the non-local measurement configuration. (b) Non-local magnetoresistance loop measured at 300 K and $V_{BG} = 0$ V. The AC current for the measurement is kept at 30 μA. (c) Hanle effect measured under out-of-plane magnetic field $B_\perp$ at various temperatures and $V_{BG} = 0$ V. The black (or gray) dots are measurement data obtained in the $P$ ($AP$) magnetic configuration. The red solid line represents the curve fitted using Eq. (1). (d) Temperature dependence of the diffusion constant $D_s$, spin relaxation time $\tau_s$, spin diffusion length $\lambda_s$, and amplitude of the non-local magnetoresistance signal $\Delta R_{NL}$.

Figure 1

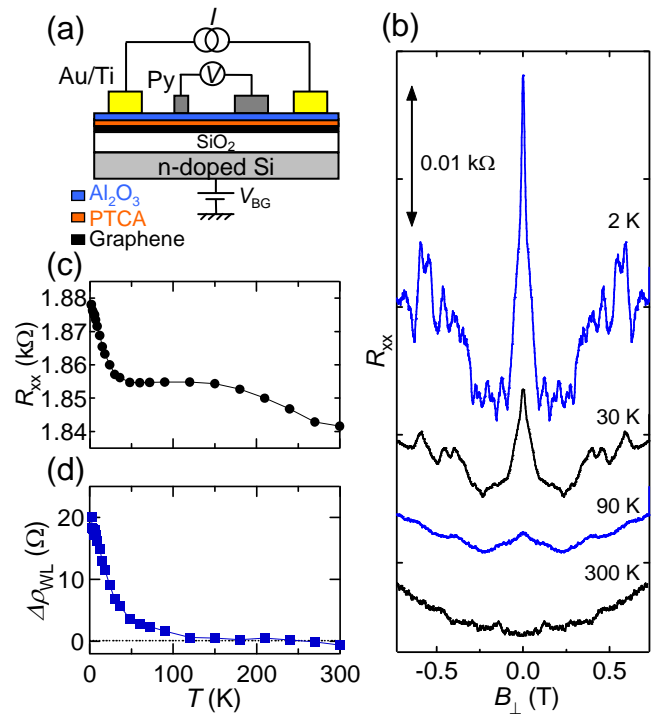

Figure 2

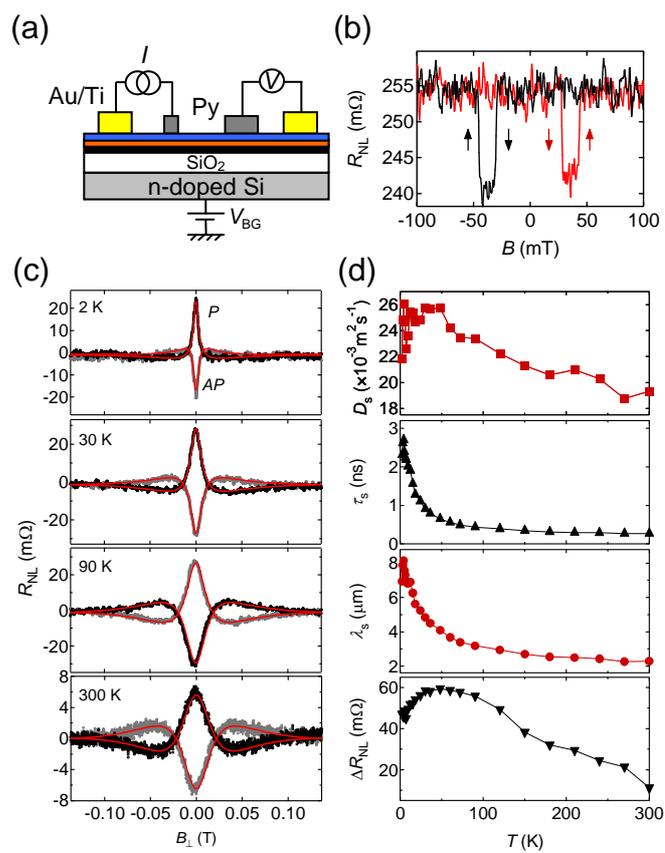